\documentclass[prd,nofootinbib,showpacs,preprint,floatfix]{revtex4}
\usepackage{latexsym}
\usepackage{dcolumn}
\RequirePackage{graphicx}

\newcommand{\MET}{\slash\!\!\!\!E_T}

\begin{document}

\preprint{ANL-HEP-PR-05-103}

\title{Angular correlations in single-top-quark and $Wjj$ production at
next-to-leading order}

\author{Zack~Sullivan}
\affiliation{High Energy Physics Division,
Argonne National Laboratory,
Argonne, Illinois 60439, USA}

\date{October 13, 2005}

\begin{abstract}
I demonstrate that the correlated angular distributions of final-state
particles in both single-top-quark production and the dominant $Wjj$
backgrounds can be reliably predicted.  Using these fully-correlated
angular distributions, I propose a set of cuts that can improve the
single-top-quark discovery significance by 25\%, and the signal to
background ratio by a factor of 3 with very little theoretical
uncertainty.  Up to a subtlety in $t$-channel single-top-quark
production, leading-order matrix elements are shown to be sufficient
to reproduce the next-to-leading order correlated distributions.
\end{abstract}


\maketitle

\section{Introduction}
\label{sec:introduction}

The discovery of single-top-quark production is one of the primary
goals of run II of the Fermilab Tevatron.  In recent years it has
become apparent \cite{CDF,D0,Acosta:2004bs,Abazov:2005zz} that a
modest $b$-tagging efficiency and larger-than-expected jet energy
resolution have promoted $Wjj$ production to the most important
background to single-top-quark production.  In order to overcome a
poor signal to background ratio of $\sim 1/10$
\cite{Acosta:2004bs,Abazov:2005zz}, recent theoretical studies
\cite{Bowen:2005rs,Cao:2004ap} have shown that only modest improvements
can be made by improving cuts in pseudorapidity or $b$-jet assignment.

It was demonstrated in Ref.\ \cite{Stelzer:1998ni} that a spin
correlation between the final-state lepton and non-$b$ jet in
single-top-quark production might provide an effective discriminate
against the $Wjj$ backgrounds.  Both the CDF and D0 Collaborations
have used this correlation at some level to improve their analyses.
Nevertheless, the theoretical basis for this has only been confirmed
at leading order (LO).  Hence, several questions arise:
\begin{enumerate}
\item Do the strong spin correlations that appear at LO in
single-top-quark production survive higher-order radiation, and leave
distinctive angular correlations in the final-state particles?

\item Is the background really insensitive to the angular distributions that
typify the signal?  If so, does this survive complex cuts on the data?

\item These correlated angular distributions are properly defined in the
reconstructed rest frame of the top quark.  How much of these correlations
is an artifact of that choice of frame?

\item Do these correlations lead to better discriminates between the
signal and backgrounds?  Are there other particle correlations that
have been missed?
\end{enumerate}

In this paper, I address all of these issues, and demonstrate the need
to account for the fully-correlated angular distributions.  First I
clarify exactly how the Mahlon-Parke \cite{Mahlon:1996pn} spin-basis
works, and why it does so surprisingly well for both $s$- and
$t$-channel single-top-quark production.  In Sec.\ \ref{sec:spincor},
I describe the exact effects of higher orders on the angular
correlations for both single-top-quark production and $Wjj$
backgrounds.  Angular correlations can be induced in the $Wjj$
background that make it look like the signal.  Avoiding these problems
requires that cuts be made on the fully-correlated (multidimensional)
angular distributions, as discussed in Sec.\ \ref{sec:corang}.  I
present some evidence that an invariant-mass distribution may be a
useful discriminate, and provide a representative set of cuts
that improve the significance by at least 25\%, and the signal to
background ratio by a factor of 3 with very little theoretical
uncertainty.

\begin{figure}[tbh]
\centering
\includegraphics[width=3.25in]{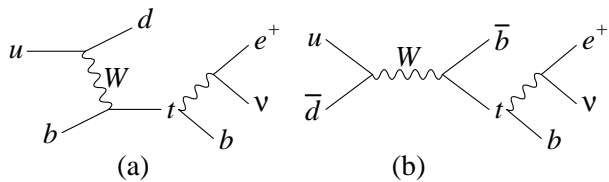}
\caption{Representative leading-order Feynman diagrams for (a) $t$-channel,
and (b) $s$-channel production of a single top quark.
\label{fig:feyn}}
\end{figure}

Before describing higher-order effects on angular distributions, we
must first understand why we expect there to be strong angular
correlations.  In Ref.\ \cite{Mahlon:1996pn}, an optimal basis was
introduced to measure the spin-induced correlations in
single-top-quark production.  The matrix elements for both $s$-channel
and $t$-channel single-top-quark production (seen in Fig.\
\ref{fig:feyn}) are proportional to
\begin{equation}
[p_d\cdot(p_t-m_t s_t)][p_e\cdot(p_t-m_t s_t)] \,,
\end{equation}
where $p_d$ and $p_e$ are the four-momenta of the down-type quark and
charged lepton in the event, $p_t$ and $m_t$ are the top-quark
four-momentum and mass, and $s_t$ is top-quark spin four-vector.  In
the top-quark rest frame $p_t = m_t (1,0,0,0)$, and $s_t = (0,
\hat{s})$.

\begin{figure}[tbh]
\centering
\includegraphics[width=2.42in]{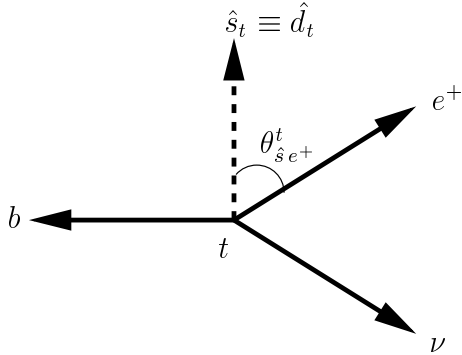}
\caption{Decay products of the top quark, and the angle
$\theta^t_{\hat{s}\,e^+}$ between the charged lepton $e^+$ and the spin
$\hat{s}_t$ of the top quark in the top-quark rest frame.  It is
convenient to choose the spin to be projected in the direction of the
down-type quark $d$ in the event.
\label{fig:tsblv}}
\end{figure}

When looking at $s$-channel production, the direction of the down-type
quark provides a convenient axis to project the top-quark spin, i.e.,
choose $\hat{s} = \hat{d}$ as in Fig.\ \ref{fig:tsblv}.  Then the
matrix element reduces to $E_d E_e m_t^2 (1+\cos \theta^t_{e^+d})$.
Since roughly 98\% of the events at the Fermilab Tevatron are produced
by pulling a $\bar d$ from the incoming antiproton, measuring $\cos
\theta^t_{e^+\bar p}$ provides the best possible measure of the spin
correlation.

Angular correlations in $t$-channel single-top-quark production are
more complicated.  The $d$ quark ends up in the highest-$E_t$
non-$b$-tagged jet $j_1$ approximately $3/4$ of the time.  Hence, it
makes sense to measure $\cos \theta^t_{e^+j_1}$.  The rest of the time
a $\bar d$-quark is in the initial state, and hence a perfect
correlation exists with the incoming hadron (mostly the antiproton at
the Tevatron).  Nevertheless, there is still a strong correlation in
these events with the direction of $j_1$.  In this case, if we project
the spin in the direction of the final-state jet (which comes from a
$\bar u$), the matrix element is proportional to $(1 +
\cos\theta^t_{d\, j_1} \cos\theta^t_{e^+j_1})$.  The dilution factor
$\cos\theta^t_{d\, j_1} = 1 - Q^2/(E^t_d E^t_{j_1})$ has a typical value
of $0.86$ with no cuts, and 95\% of the time is more than
$0.5$.  The dilution is a small effect in a small fraction of the
events.  Hence, $\cos \theta^t_{e^+j_1}$ is an excellent quantity to
measure.  The saving grace here was a kinematic correlation induced by
the $t$-channel exchange of the $W$ boson.  We will return to the
effects of these ``kinematically-induced correlations'' in
single-top-quark and $Wjj$ production in each of the Sections below.

\section{Angular correlations at LO and NLO}
\label{sec:spincor}

Given the strength of the spin correlations in single-top-quark
production, it is natural to want to use these as a discriminate
between the signal and backgrounds.  In order to be effective,
however, we must know that these correlations will actually appear in
the data.  Before reaching the issue of detector effects, we must
first address whether these correlations are an artifact of the
leading-order calculations that found them, or are real.  We begin
with the dominant $t$-channel production cross section.

There are only three ways that QCD radiation at next-to-leading order
can degrade the measured angle between the lepton and jet $j_1$ in
$t$-channel production.  The first is radiation off of the top quark
before it decays.  This can cause a spin flip, but is suppressed by
the top quark mass.  The second way to dilute the correlation is for
the down-type quark to radiate, and thereby change the measured angle.
This is suppressed for typical jets because most radiation is fairly
soft and collinear, and is reabsorbed into the final measured jet.
Only very energetic wide-angle jets are relevant.  The most
significant dilution of the spin correlation should therefore come
from misidentifying the jet that includes the down-type quark.  From
this point of view, the effect of next-to-leading order (NLO)
radiation is to provide additional jets to misidentify.

The efficiency of $\cos \theta^t_{ej_1}$ as a discriminate for the
single-top-quark signal, will ultimately be based upon experimental
precision.  Nevertheless, we can calculate whether there us an
underlying limit based on the rate of misidentification of the
direction of the down-type quark in the event.  For all calculations
in this paper I use MCFM 4.1 \cite{MCFM} (with some
corrections\footnote{These have been provided to, and confirmed with,
the authors of MCFM.} to the matrix element for $t$-channel
single-top-quark production based on the NLO code ZTOP
\cite{Sullivan:2004ie}).  MCFM contains NLO corrections to top-quark
decay, as well as full spin correlations for final-state leptons and
jets.  Figures are presented for $t$ (not $\bar t$) production only at
run II of the Fermilab Tevatron, a $1.96$ TeV $p\bar p$ collider, but
the underlying principles apply equally well to the CERN Large Hadron
Collider (LHC).  CTEQ6L1 and CTEQ6M \cite{Pumplin:2002vw,Stump:2003yu}
parton distributions are used for LO and NLO distributions,
respectively.  The top-quark mass is taken to be 175 GeV.

Acceptance cuts for inclusive $W+2$-jet distributions are based on
Ref.\ \cite{Acosta:2004bs} and listed in Table \ref{tab:gencuts}.  The
top quark is reconstructed using the ``$b$-jet'' (chosen randomly if
there are an even number of candidates in the final state) and a
reconstructed $W$ boson of mass $80.4$ GeV.  The $W$ boson is
reconstructed using an isolated charged lepton and missing transverse
energy $\MET$, where the neutrino solution with the smallest absolute
pseudorapidity is chosen.  Other neutrino solutions were tested, but
found to give worse fits to the top-quark mass.  A loose cut on the
top-quark mass $M_{be\,\MET}$ has little effect on the shapes of the
distributions presented, but does change the normalization of the
backgrounds.

\begin{table}[tbh]
\caption{Acceptance cuts applied to $E_T$-ordered jets and leptons in the
$Wjj$ final state.  Require one charged lepton (denoted $e$
throughout), missing transverse energy $\MET$, and at
least two jets.  At least one jet must pass the $b$ acceptance (two
$b$ jets for $s$-channel single-top-quark production).
\label{tab:gencuts}}
\begin{center}
\begin{ruledtabular}
\begin{tabular}{l}
$E_{Tj} > 15$ GeV, $|\eta_j| < 2.8$, $\Delta R_{k_T} < 0.54$ ($\approx
\Delta R_\mathrm{cone} < 0.4$)\\
$E_{Tb} > 25$ GeV, $|\eta_b| < 1.4$\\
$E_{Te} > 15$ GeV, $|\eta_e| < 1.4$, $\Delta R_{ej} > 0.4$\\
$\MET > 15$ GeV\\
$140 < M_{be\,\MET} < 210$
\end{tabular}
\end{ruledtabular}
\end{center}
\end{table}

\begin{figure}[tbh]
\centering
\includegraphics[width=3.25in]{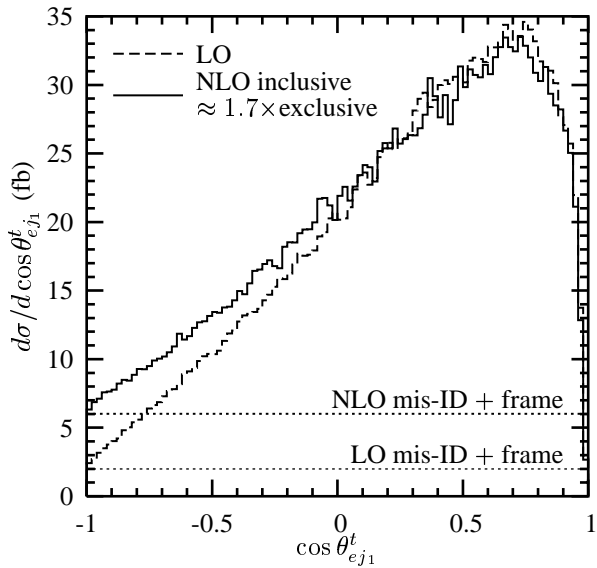}
\caption{Cosine of angle ($\cos \theta^t_{ej_1}$) between the charged
lepton and highest-$E_T$ light-quark jet in the top-quark rest frame
of $t$-channel single-top-quark production at LO and NLO.  The lepton
isolation cut suppresses events at large $\cos \theta^t_{ej_1}$.  A
nearly angle-independent underlying contribution comes from
misidentification of which jet contains the down-type quark.
\label{fig:costejo}}
\end{figure}

Let us compare $t$-channel production at LO and NLO.  In Fig.\
\ref{fig:costejo} we see the differential cross section as a function
of $\cos \theta^t_{ej_1}$ at LO.  For the cuts in Table
\ref{tab:gencuts}, approximately $4.5\%$ of the events come from a
$\bar d$, $\bar s$, or $\bar b$ quark in the initial state.  The
correct direction to project into for these events is the proton
direction, since most $\bar d$ come from the antiproton.  All of the
other events (except the small sea-quark contribution) should be in
the direction of the highest-$E_T$ jet that did not come from the
top-quark decay.  So we expect to see a $100\%$ correlation on top of
an uncorrelated\footnote{There is a kinematic anti-correlation between
the direction of the proton and the jet containing the $\bar u$ or
$\bar c$ that produces a slight tilt to the background.} 
``background'' of about $4.5\%$.  The dip near $\cos
\theta^t_{ej_1}=1$ is due to the lepton isolation cut.

When we place cuts at LO, the forward jets and leptons are excluded.
Therefore, an additional dilution factor of $0.95$ is added to the
case where the the down-type quark was in the initial state.  In Fig.\
\ref{fig:costejo} we are reconstructing the top-quark mass by fitting
the missing transverse energy $\MET$ to the $W$-boson mass.  This adds
a dilution factor of $0.95$ to all production modes.  The final cross
section is then
\begin{equation}
\frac{d\sigma}{d\cos \theta^t_{ej_1}} \approx 21[ 0.75(1+ 0.95\cos
\theta^t_{ej_1}) + 0.25(1+0.95\times0.95\times0.86\cos
\theta^t_{ej_1})] \, \mbox{fb} \,.
\end{equation}

At NLO the dilution factor from frame reconstruction is slightly worse
at $0.9$, but the main difference is that there are more jets to
misidentify as containing the down-type quark.  It was demonstrated in
Refs.\ \cite{Sullivan:2004ie,Stelzer:1998ni} that a sizable fraction of
the leading jets $j_1$ are actually from wide-angle emission of
initial-state $\bar b$ jets.  For the cuts of Table \ref{tab:gencuts},
about $8.3\%$ of the events contain a $\bar b$ quark.  This particle
is completely uncorrelated with the direction of the charged lepton.
If we add these two effects, we find the correct uncorrelated
``background'' underneath the signal.  The conclusion to be drawn is
that the spin correlations are robust when higher-order radiation is
included, but confusion in picking the correct correlated jet will
slightly dilute the signal in a completely calculable way.

\begin{figure}[tbh]
\centering
\includegraphics[width=3.25in]{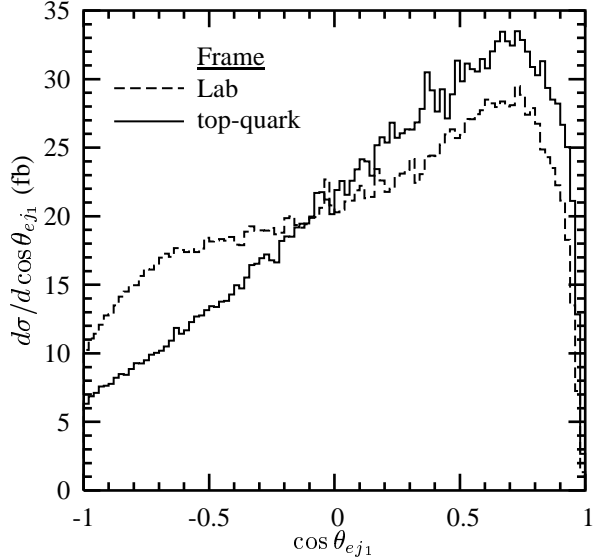}
\caption{Cosine of angle ($\cos \theta_{ej_1}$) between the charged
lepton and highest-$E_T$ light-quark jet $j_1$ in the lab frame and
top-quark rest frame of single-top-quark production at NLO.
\label{fig:cosltejo}}
\end{figure}

We can appreciate the stability of the spin correlations at NLO by
comparing the correlation in the top-quark rest frame to the
laboratory frame.  We see in Fig.\ \ref{fig:cosltejo} that the lab
frame is only slightly worse for observing the correlation.  This is
due to the fact the top-quark is typically nonrelativistic, and so
there is not much of a boost in switching frames.  If the top quark
momentum were comparable to its mass, the cosine of the angle in the
lab frame $\cos \theta^l_{ej_1}$ would be flat.  This leads to the
fortunate accident that the angular correlation in $t$-channel
production is not very sensitive to how well the top-quark rest-frame
is reconstructed.

\begin{figure}[tbh]
\centering
\includegraphics[width=3.25in]{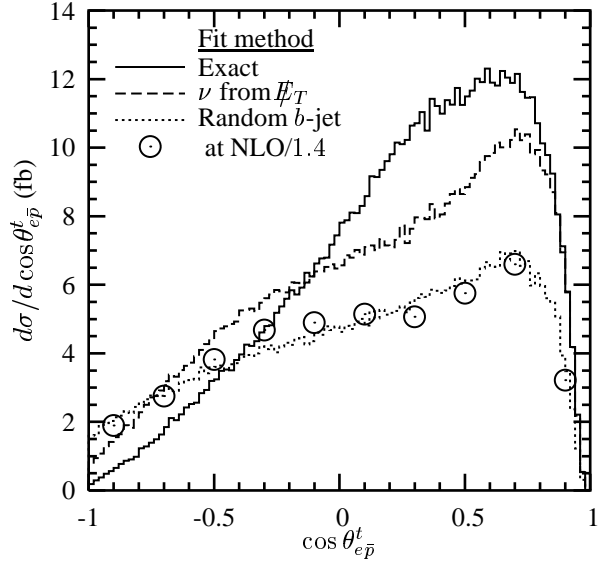}
\caption{Cosine of the angle ($\cos \theta^t_{e\bar p}$) between the charged
lepton and antiproton in the top-quark rest frame of $s$-channel
single-top-quark production at LO at the Fermilab Tevatron (a $1.96$
TeV $p\bar p$ collider).  The solid line corresponds to reconstructing
the top-quark frame using the exact neutrino and $b$-jet from the
top-quark decay.  The dashed line uses the missing transverse energy
and $W$ mass to fit a putative neutrino momentum.  The dotted line
adds the effect of using a randomly chosen $b$-jet.  The rescaled NLO
cross section with a randomly chosen $b$ jet is indicated by open
circles.
\label{fig:costepb}}
\end{figure}

In $s$-channel production, the direction of the charged lepton and the
antiproton are well measured.  The limiting factor in utilizing $\cos
\theta^t_{e\bar p}$ is how well the top-quark rest-frame is reconstructed.
Since the neutrino coming from top-quark decay is not observed, it is
typically reconstructed from the missing transverse energy and a fit
to the $W$ mass.  We see in the dashed line of Fig.\ \ref{fig:costepb}
the already sizable effect of this approximate fit on the distribution
of $\cos \theta^t_{e\bar p}$.  $s$-channel production has the
additional difficulty that it is typically impossible to tell which of
the two $b$-jets came from the top quark decay.  The simplest choice
for reconstructing the top quark is to randomly choose one of the two
$b$ jets.  The dotted line in Fig.\ \ref{fig:costepb} shows what the
measured correlation would look like given this choice.  We will
investigate in Sec.\ \ref{sec:corang} whether other choices of
assigning the $b$ jet that came from the top quark are useful
discriminates from the background.  It is apparent, however, that the
shape of this distribution is the same at LO and NLO.  The only
difference is a $K$-factor of $1.4$, which is consistent with the
results of Refs.\ \cite{Harris:2002md,Sullivan:2004ie}.

\begin{figure}[tbh]
\centering
\includegraphics[width=3.25in]{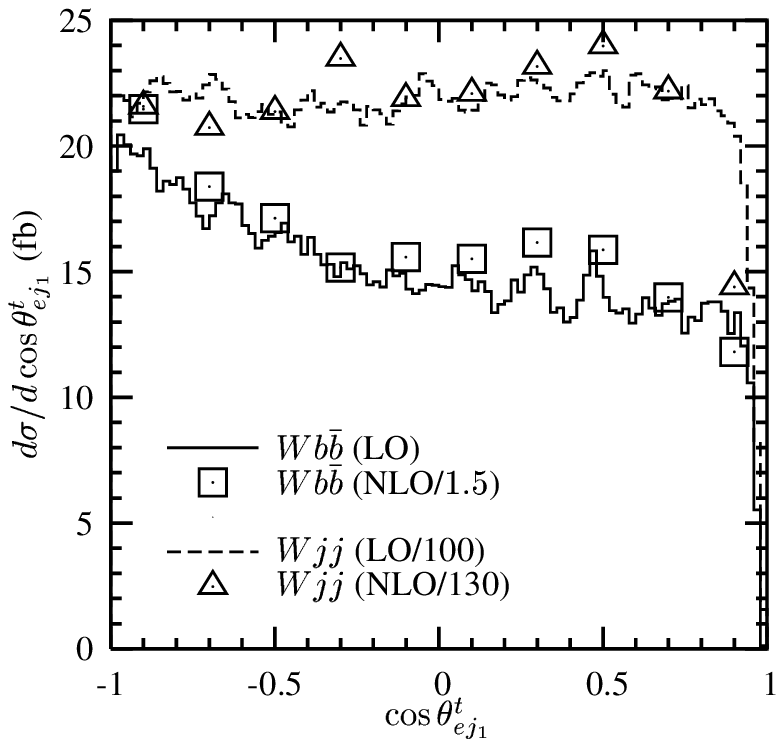}
\caption{Cosine of angle ($\cos \theta^t_{ej_1}$) between the charged
lepton and highest-$E_T$ light-quark jet $j_1$ in the $M_{Wb}$ rest
frame of $Wb\bar b$ (solid), and $Wjj$ (dashed) production at LO,
where one of the two highest-$E_T$ jets is randomly tagged as a
$b$-jet, and the other is $j_1$.  The NLO cross sections are shown
scaled to the LO cross sections.
\label{fig:wcostejo}}
\end{figure}

\begin{figure}[tbh]
\centering
\includegraphics[width=3.25in]{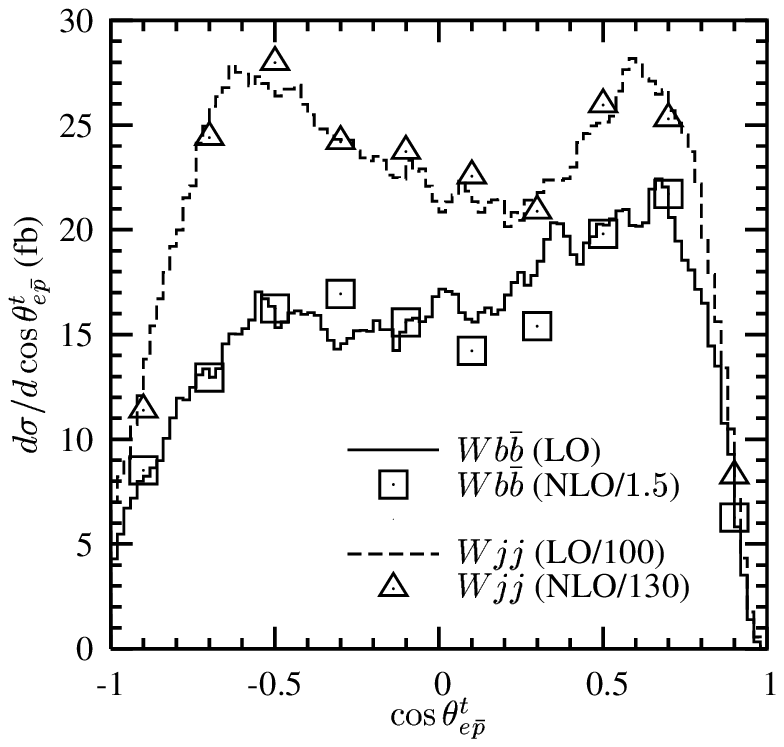}
\caption{Cosine of angle ($\cos \theta^t_{e\bar p}$) between the charged
lepton and antiproton in the $M_{Wb}$ rest frame of $Wb\bar b$
(solid), and $Wjj$ (dashed) production at LO, where one of the two
highest-$E_T$ jets is randomly tagged as a $b$-jet.  The NLO cross
sections are shown scaled to the LO cross sections.
\label{fig:wcostepb}}
\end{figure}

It was shown in Refs.\ \cite{Stelzer:1998ni,O'Neil:2002ks} that the
$Wjj$ background at LO is almost flat in the distribution $\cos
\theta^t_{ej_1}$.  We now wish to test this at NLO.  In Fig.\
\ref{fig:wcostejo} the $Wjj$ and $Wb\bar b$ backgrounds\footnote{Note,
the QCD $Wc\bar c$ and $Wb\bar b$ backgrounds are identical before
detector effects are applied.} are shown at LO and NLO scaled to match
the LO cross section.  It appears that LO provides a good description
of the shape of the $\cos \theta^t_{ej_1}$ distribution.  The shape of
the $\cos \theta^t_{e\bar p}$ distribution is shown in Fig.\
\ref{fig:wcostepb}, and is also well-approximated by LO times a NLO
$K$-factor.  Therefore, up to the issues mentioned in $t$-channel
reconstruction, LO matrix elements provide excellent approximations to
both the signal and backgrounds for single-top-quark production.

\section{Correlated angular distributions}
\label{sec:corang}

\begin{figure}[tbh]
\centering
\includegraphics[width=3.25in]{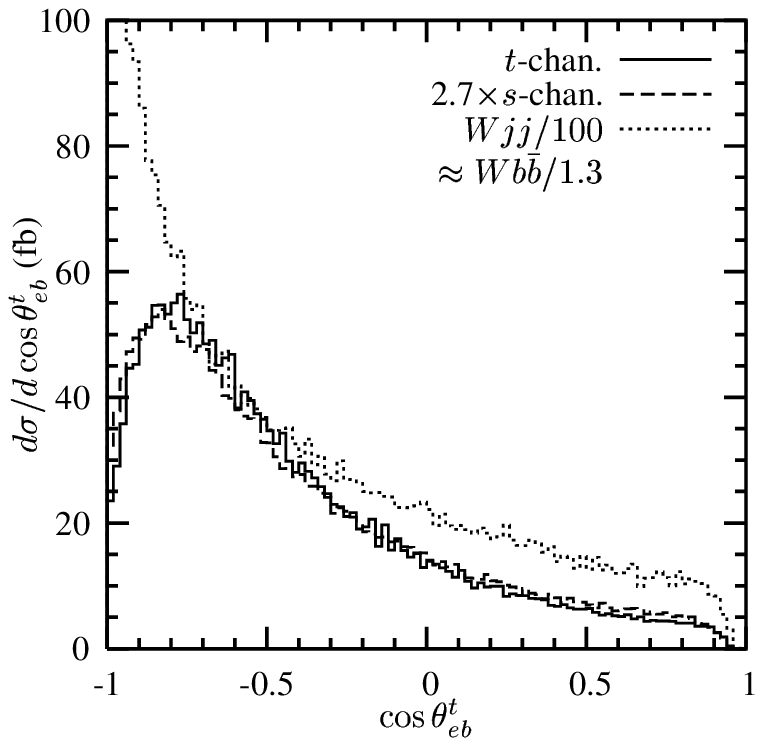}
\caption{Cosine of the angle ($\cos \theta^t_{eb}$) between the
charged lepton and final-state $b$ quark in the top-quark rest frame.
The $t$-channel and $s$-channel single-to-quark distributions are the
same, since they both involve a real top-quark decay.  The $Wjj$ and
$Wb\bar b$ distributions appear similar to real top-quark decay
because of the boost into the top-quark rest frame.
\label{fig:cteb}}
\end{figure}

\begin{figure}[tbh]
\centering
\includegraphics[width=2.26in]{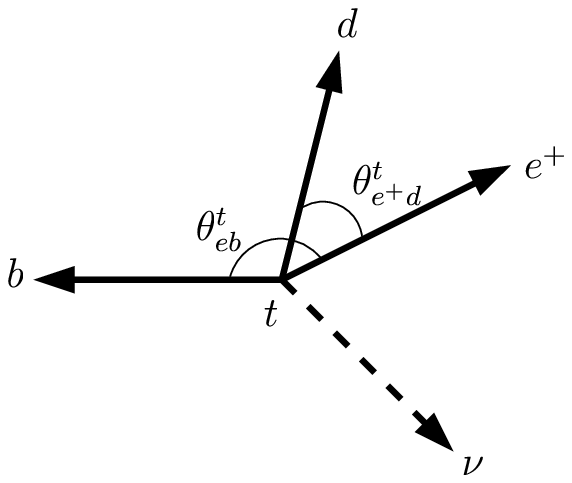}
\caption{Particles in the final state of single-top-quark production,
in the top-quark rest frame.  The angle between the charged lepton
$e^+$ and the $b$ quark is expected to be larger than the angle
between $e^+$ and the down-type quark $d$.
\label{fig:tbdlv}}
\end{figure}

Having established that the angular correlations are stable between LO
and NLO, we now turn to whether they are a useful discriminant.  The
first correlation we examine is between the charged lepton, denoted as
$e^+$, and the $b$ jet from the top-quark decay.  In a top-quark decay
the $b$ jet recoils against a real $W$, which then decays to two
leptons.  We therefore expect there to be a large angle between $e^+$
and $b$.  This is borne out in the distributions shown in Fig.\
\ref{fig:cteb}.  Recall that for $t$-channel production the
charged lepton and $j_1$ have a $1+\cos\theta^t_{ej_1}$ distribution.
We expect, then, that $\cos\theta^t_{eb} < \cos\theta^t_{ej_1}$, as
depicted in Fig.\ \ref{fig:tbdlv}.  This is true roughly $80\%$ of the
time for both $s$-channel and $t$-channel production.  Hence, we might
expect we have a better discriminate than $b$-tagging for $t$-channel
production, and an excellent way to determine which of the two
$b$-jets in $s$-channel production came from the top-quark decay.

\begin{figure}[tbh]
\centering
\includegraphics[width=3.25in]{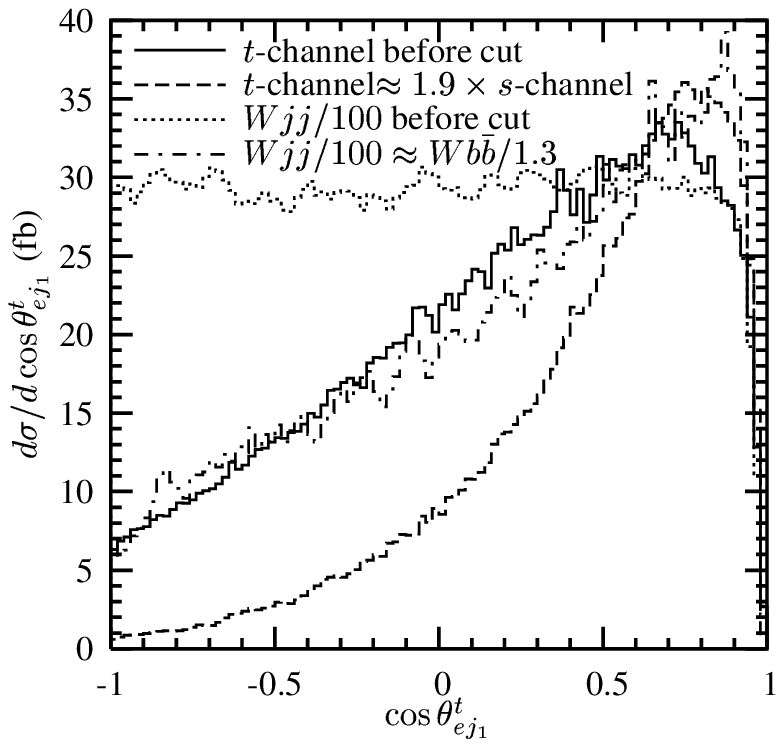}
\caption{Cosine of the angle ($\cos \theta^t_{ej_1}$) between the
charged lepton and final-state jet $j_1$ in the top-quark rest frame,
before and after the cut $\cos \theta^t_{eb} < \cos \theta^t_{ej_1}$.
The $t$-channel and $s$-channel single-to-quark distributions are the
same after cuts.  The $Wjj$ and $Wb\bar b$ distributions go from
approximately flat to very similar to the $t$-channel distribution
before cuts.
\label{fig:ctejcut}}
\end{figure}

Ignoring $b$-tagging, let us begin with the two highest-$E_T$ jets in
the event.  Define the ``$b$-jet'' to be the jet with the largest
opening angle in the reconstructed $e$-jet-$\MET$ rest
frame (the putative top-quark rest frame before mass cuts).  
Up to a normalization factor of $1.3$, the $\cos \theta^t_{e\bar p}$
for $s$-channel production now follows the exact result in Fig.\
\ref{fig:costepb} when $\cos \theta^t_{e\bar p}<-0.2$, and the dashed
line from Fig.\ \ref{fig:costepb} at larger $\cos \theta^t_{e\bar
p}$.  Hence, this definition effectively removes the $b$-jet
assignment uncertainty in the accepted events.  The real question is:
what does this do to the $Wjj$ background in general?  In Fig.\
\ref{fig:ctejcut} we see that it takes the initially flat $Wjj$
backgrounds, and shapes them to look just like the original
$t$-channel signal!  It looks like we've taken a step backward, but we
will see in Sec.\ \ref{sec:otherang} that this will be fine.

\begin{figure}[tbh]
\centering
\includegraphics[width=1.66in]{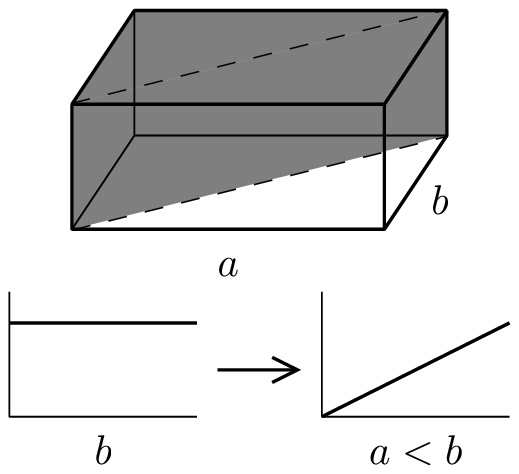}\hspace*{.03in}\includegraphics[width=1.56in]{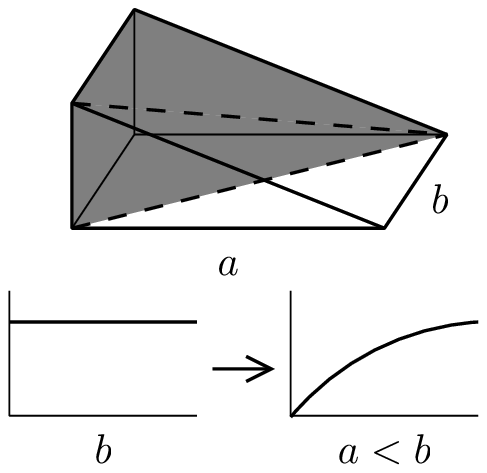}
\caption{Effect of the cut $a<b$ on an initially flat distribution $b$ for
two cases: (left) completely flat distributions, (right) a falling
distribution for $a$ (similar to the actual $\cos\theta^t_{eb}$).
The resulting shape for $b$ is an integral over $a$, and is only
modestly sensitive to the detailed shape of $a$.
\label{fig:fakecor}}
\end{figure}

First, let us understand why the $Wjj$ distribution has the shape it
does.  This is a generic effect of the cut we have made.  On the left
side of Fig.\ \ref{fig:fakecor} we see a function that is completely
flat in two variables $a$ and $b$.  If we make the cut $a<b$, and plot
the projection of $b$, we immediately see that we have induced a
correlation that was not previously present.  The right side of Fig.\
\ref{fig:fakecor} is closer to the situation we have from
Fig.\ \ref{fig:cteb}, with $a\approx \cos\theta^t_{eb}$ and $b\approx
\cos\theta^t_{ej_1}$ for $Wjj$ (or $Wb\bar b$).  Aside from a small variation,
the rough shape is always the same.  The reason $\cos\theta^t_{eb}$ is
peaked at large angles is that we have shifted the distribution, which
is symmetric in the lab frame, by boosting the top-quark rest frame
defined by the $W$ and $b$.  This should serve as a reminder that any
cut on angular distributions will induce some correlation.

Though we may be tempted to ignore this angular cut as a failed trial,
we should recognize that there are two common experimental realities
that are roughly equivalent to the cut $\cos \theta^t_{eb} <
\cos \theta^t_{ej_1}$.  First, the efficiency for tagging (or
mistagging) a jet generally increases with its transverse energy.
Since the highest-$E_T$ jet tends to balance the $W$ boson, its angle
with the lepton will on average be larger than for the non-tagged jet.
The second common occurrence is the use of a loose top-quark mass
reconstruction.  Even if the tagging rate was energy-independent, the
top quark mass is large enough that only a tagged jet that recoils
against the $W$ will pass the cut.  Both cases preferentially choose
the jet with the largest opening angle to be the tagged jet.  Hence,
this kinematic correlation will likely always be present at some level
in a realistic analysis.

\subsection{Using full angular correlations}
\label{sec:otherang}

\begin{figure}[tbh]
\centering
\includegraphics[width=3in]{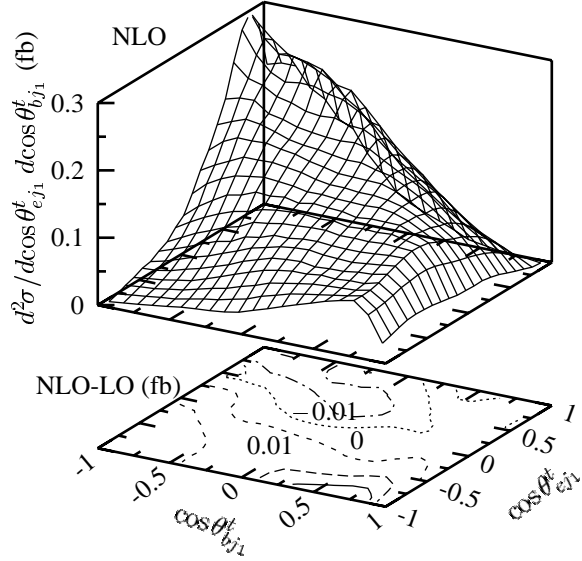}
\caption{Correlated angular distributions of the final-state particles
in the top-quark rest frame of $t$-channel single-top-quark production
at NLO (top), and the difference between NLO and LO (bottom).  This is
a two-dimensional projection between $\cos\theta^t_{ej_1}$, where $e$
is the charged lepton and $j_1$ is tagged as the highest-$E_T$
light-quark jet, and $\cos\theta^t_{bj_1}$, where $b$ is the $b$ jet
from the top-quark decay.
\label{fig:tteobo}}
\end{figure}

\begin{figure}[tbh]
\centering
\includegraphics[width=3in]{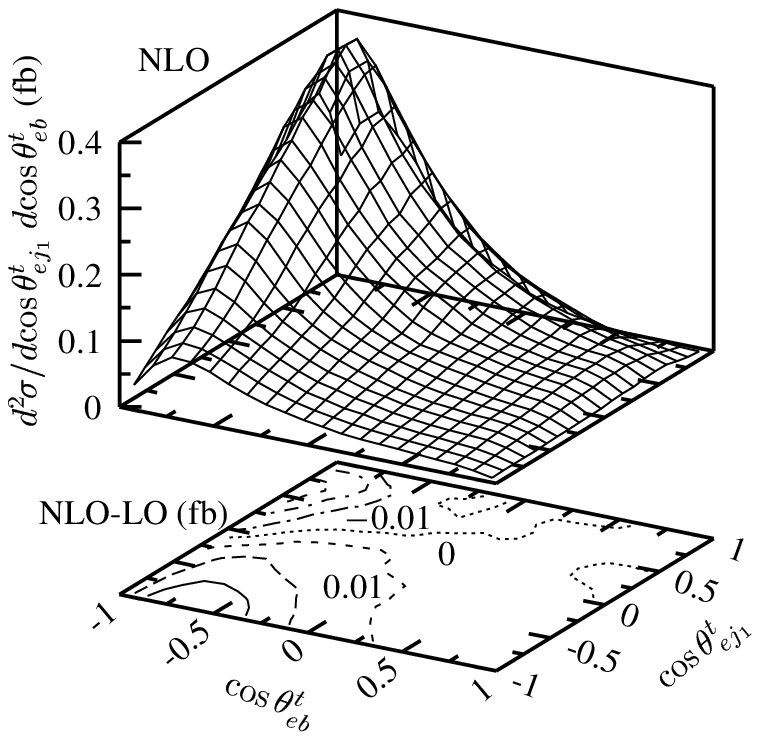}
\caption{Correlated angular distributions of the final-state particles
in the top-quark rest frame of $t$-channel single-top-quark production
at NLO (top), and the difference between NLO and LO (bottom).  
Same as Fig.\ \protect\ref{fig:tteobo}, but projecting on
$\cos\theta^t_{ej_1}$ and $\cos\theta^t_{eb}$.
\label{fig:tteoeb}}
\end{figure}

\begin{figure}[tbh]
\centering
\includegraphics[width=3in]{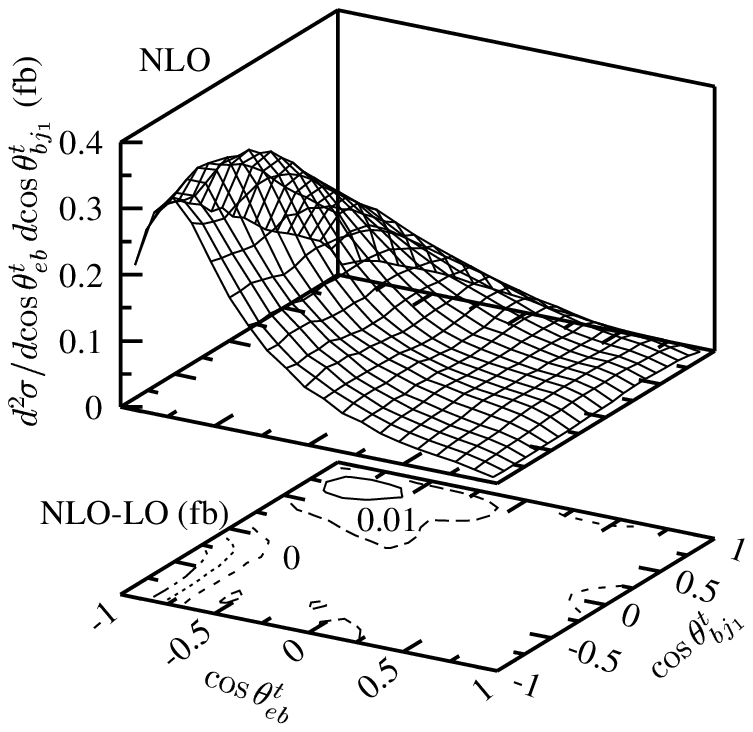}
\caption{Correlated angular distributions of the final-state particles
in the top-quark rest frame of $t$-channel single-top-quark production
at NLO (top), and the difference between NLO and LO (bottom).  
Same as Fig.\ \protect\ref{fig:tteobo}, but projecting on
$\cos\theta^t_{eb}$ and $\cos\theta^t_{bj_1}$.
\label{fig:ttebbo}}
\end{figure}

\begin{figure}[tbh]
\centering
\includegraphics[width=3.125in]{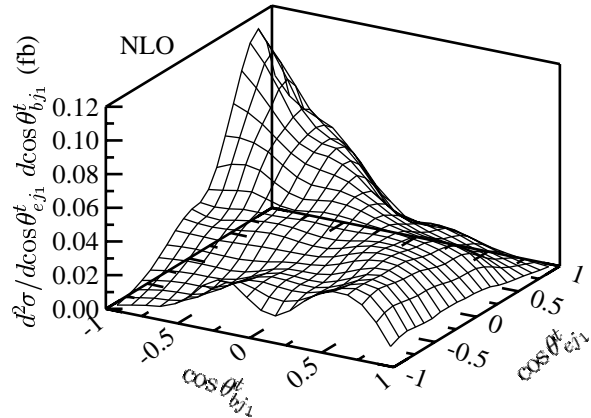}
\caption{Correlated angular distributions of the final-state particles
in the top-quark rest frame of $s$-channel single-top-quark production
at NLO.
Labels are the same as in Fig.\ \protect\ref{fig:tteobo}.
\label{fig:tseobo}}
\end{figure}

\begin{figure}[tbh]
\centering
\includegraphics[width=3.125in]{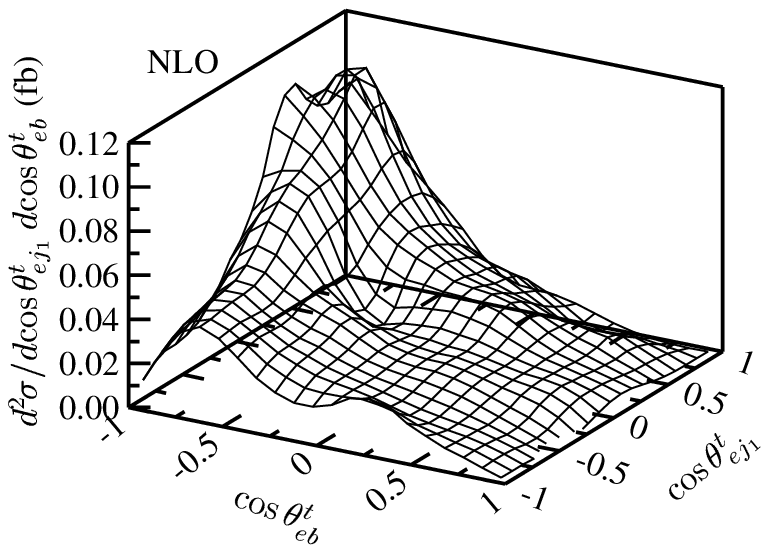}
\caption{Correlated angular distributions of the final-state particles
in the top-quark rest frame of $s$-channel single-top-quark production
at NLO.
Same as Fig.\ \protect\ref{fig:tseobo}, but projecting on
$\cos\theta^t_{ej_1}$ and $\cos\theta^t_{eb}$.
\label{fig:tseoeb}}
\end{figure}

\begin{figure}[tbh]
\centering
\includegraphics[width=3.125in]{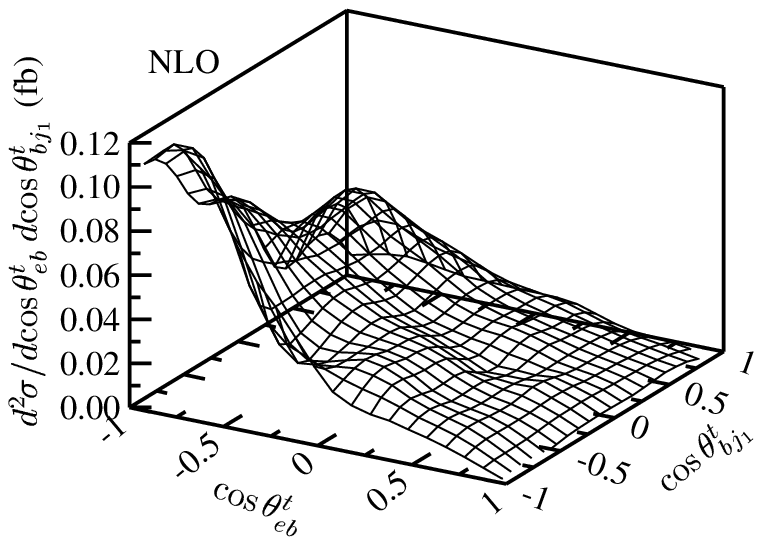}
\caption{Correlated angular distributions of the final-state particles
in the top-quark rest frame of $s$-channel single-top-quark production
at NLO.
Same as Fig.\ \protect\ref{fig:tseobo}, but projecting on
$\cos\theta^t_{eb}$ and $\cos\theta^t_{bj_1}$.
\label{fig:tsebbo}}
\end{figure}

If we wish to make cuts on angular distributions, we must take into
account the full angular correlations.  There are at least three observed
objects in the final state: the charged lepton $e$ (or $\mu$), the $b$
from the top-quark decay, and the other jet $j_1$.  In Figs.\
\ref{fig:tteobo}--\ref{fig:ttebbo} we plot the correlated NLO distributions
between pairs of these angles in $t$-channel single-top-quark production.
Also shown is the difference (in fb) between the NLO and LO distributions.
Except for where new phase space opens up, the difference is less than 3\%.
The same distributions are shown for $s$-channel production in Figs.\
\ref{fig:tseobo}--\ref{fig:tsebbo}.  The difference between NLO and LO
times a $K$-factor of $1.43$ is too small to display.

\begin{figure}[tbh]
\centering
\includegraphics[width=3.125in]{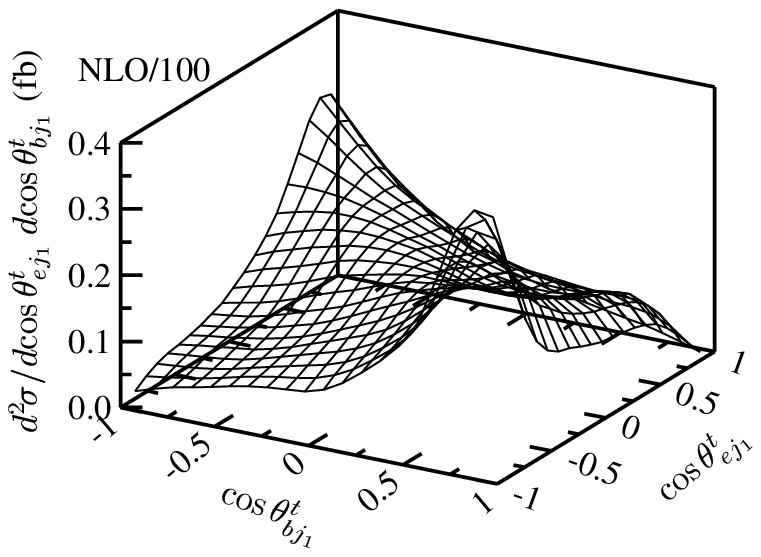}
\caption{Correlated angular distributions of the final-state particles
in the reconstructed top-quark rest frame of $Wjj$ production at NLO.
Labels are the same as in Fig.\ \protect\ref{fig:tteobo}.
\label{fig:twjjeobo}}
\end{figure}

\begin{figure}[tbh]
\centering
\includegraphics[width=3.125in]{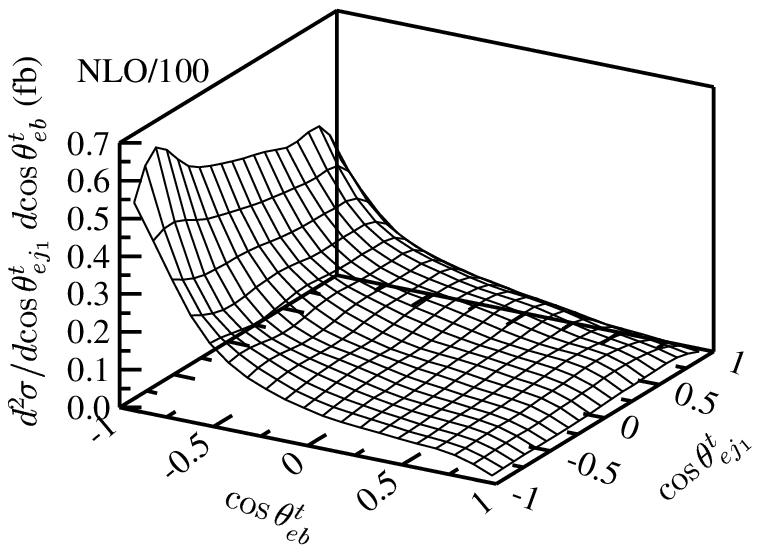}
\caption{Correlated angular distributions of the final-state particles
in the reconstructed top-quark rest frame of $Wjj$ production at NLO.
Same as Fig.\ \protect\ref{fig:twjjeobo}, but projecting on
$\cos\theta^t_{ej_1}$ and $\cos\theta^t_{eb}$.
\label{fig:twjjeoeb}}
\end{figure}

\begin{figure}[tbh]
\centering
\includegraphics[width=3.125in]{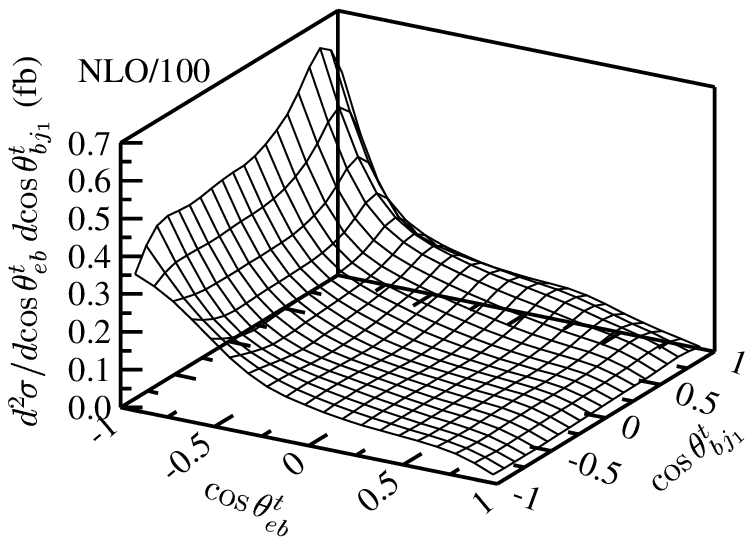}
\caption{Correlated angular distributions of the final-state particles
in the reconstructed top-quark rest frame of $Wjj$ production at NLO.
Same as Fig.\ \protect\ref{fig:twjjeobo}, but projecting on
$\cos\theta^t_{eb}$ and $\cos\theta^t_{bj_1}$.
\label{fig:twjjebbo}}
\end{figure}

We can see in Figs.\ \ref{fig:tteoeb} and \ref{fig:tseoeb} that the
cut we examined in Sec.\ \ref{sec:corang} was sensible for both
$t$-channel and $s$-channel production as most of the events are piled
up at small $\cos\theta^t_{eb}$ and large $\cos\theta^t_{ej_1}$.  A slightly
better cut would follow the contours of the peak, but this depends on
the $Wjj$ background that we show in Figs.\
\ref{fig:twjjeobo}--\ref{fig:twjjebbo}.  The first thing to notice is
that our apparent shaping of the background actually was an artifact
of removing one peak in the correlated distribution.  Given the
symmetry of the jets, the initial flat background for the
$\cos\theta^t_{ej_1}$ distribution was due to the sum over two peaks
in the full phase space with broad tails that accidentally compensated
each other.

Based on the correlated distributions for the signal and background
presented in Figs.\ \ref{fig:tteobo}--\ref{fig:twjjebbo}, I propose
the following series of acceptance cuts as a starting point to better
isolate single-top-quark production:
\begin{enumerate}
\item $\cos\theta^t_{eb} < \cos\theta^t_{ej_1}$.
\item $\cos\theta^t_{bj_1} < \cos\theta^t_{ej_1}$.
\item $\cos\theta^t_{bj_1} < 0.6$--$0.8$.
\item $\cos\theta^t_{ej_1} > 0$--$0.4$ or $\cos\theta^t_{eb} > -0.8$.
\end{enumerate}
These cuts should be optimized based on full detector simulations, and
replaced by more sophisticated parameterizations of the correlated
space.  We saw in Sec.\ \ref{sec:spincor}, and the $t$-channel
plots\footnote{Differences between NLO and LO times a $K$ factor in
$s$-channel and $Wjj$ correlated distributions are too small to
reliably calculate, but are typically much less than 5\%.} above, that
a leading-order matrix is good enough to represent the NLO
correlations.  Therefore, feeding the correlated matrix elements into
a showering Monte Carlo like PYTHIA \cite{Sjostrand:2000wi} or HERWIG
\cite{Corcella:2000bw} should allow for accurate modeling of the
acceptances.

\begin{figure}[tbh]
\centering
\includegraphics[width=3.25in]{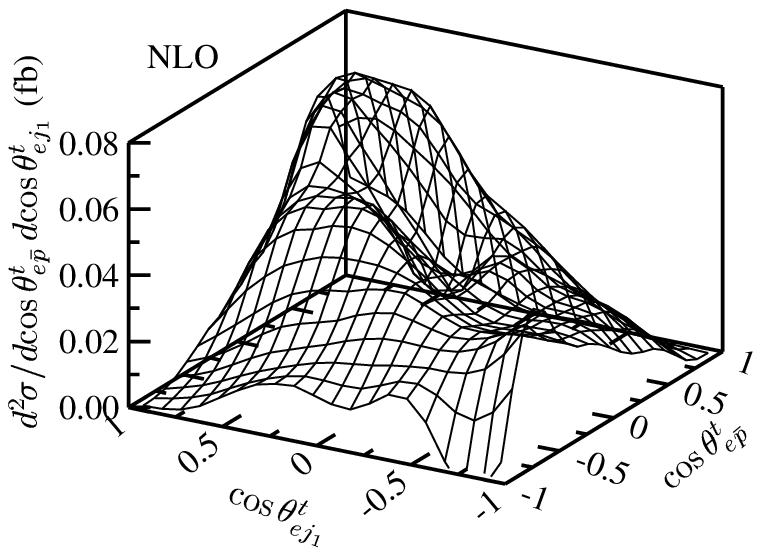}
\caption{Correlated angular distributions of the final-state particles
in the top-quark rest frame of $s$-channel single-top-quark production
at NLO.
Same as Fig.\ \protect\ref{fig:tseobo}, but projecting on
$\cos\theta^t_{e\bar p}$ and $\cos\theta^t_{ej_1}$, where $\bar p$ is the
incoming antiproton.
\label{fig:tsepeo}}
\end{figure}

\begin{figure}[tbh]
\centering
\includegraphics[width=3.25in]{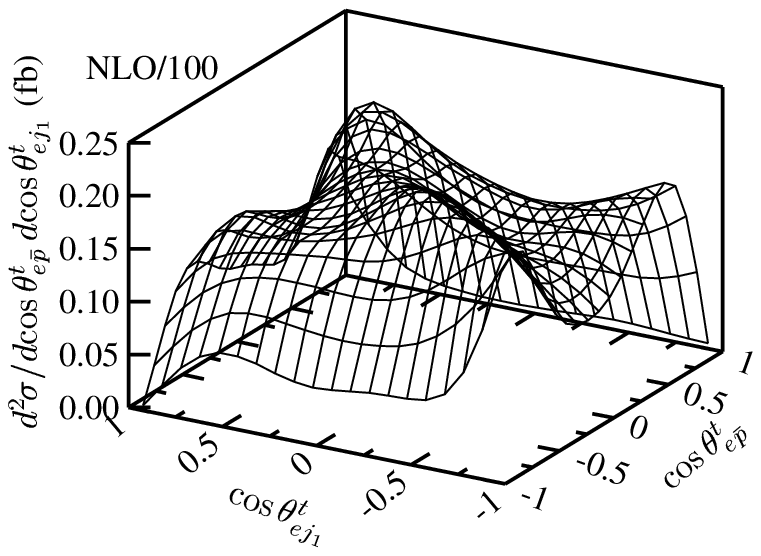}
\caption{Correlated angular distributions of the final-state particles
in the reconstructed top-quark rest frame of $Wjj$ production at NLO.
Labels are the same as in Fig.\ \protect\ref{fig:tsepeo}.
\label{fig:twjjepeo}}
\end{figure}

The $s$-channel production mode can make use of one strong additional
acceptance cut.  Most of the signal is contained in the box defined by
$\cos\theta^t_{e\bar p} > -0.2$ and $\cos\theta^t_{ej_1} > 0.1$, as
seen in Fig.\ \ref{fig:tsepeo}.  The $Wjj$ backgrounds in Fig.\
\ref{fig:twjjepeo}, however, are nearly flat in both of these
variables, except at the edges of phase space where jet cuts come in.
While I do not include this cut in the numerical results below, up to
$3/4$ of the background may be removed for a small loss in signal if
this cut is layered on top of the cuts listed above.

\subsection{An additional discriminate}
\label{sec:masscut}

Now that fully differential NLO spin-dependent calculations are available,
we can look for other correlations to separate signals and
backgrounds.  Subtle effects, such as how to define the neutrino in
$W$ reconstruction can have a sizable impact on measured cross
sections.  One of the other limiting issues in applying a top-quark
mass cut is exactly how well the missing energy can be measured, and
fed into the top-quark reconstruction.  To attempt to avoid this
problem, I have also looked at the fully correlated combinations of
invariant masses $M_{ij}$ at NLO after cuts.

\begin{figure}[tbh]
\centering
\includegraphics[width=3.25in]{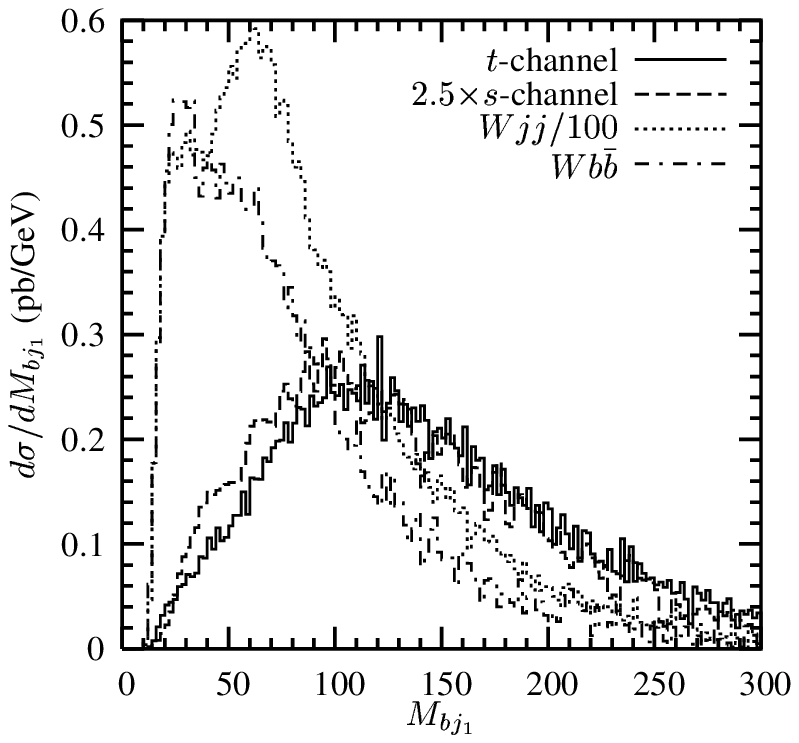}
\caption{Invariant mass of the identified $b$-jet and highest-$E_T$ jet
$j_1$ from $t$- and $s$-channel single-top-quark, $Wjj$, and $Wb\bar b$
production.
\label{fig:mbjo}}
\end{figure}

Most of the multi-dimensional correlated $M_{ij}$ plots are similar
between the single-top-quark signals and the $Wjj$ backgrounds.
Specifically, the tails of the distributions are almost identical, and
the peak for signal is under a strongly rising background.  Some cuts
may help, but it is difficult to retain enough signal.  The one
exception is the projection along $M_{bj_1}$, as shown in Fig.\
\ref{fig:mbjo}.  Applying a cut of $M_{bj_1}>50$--$120$ GeV can reduce
the background whether or not the top-quark mass can be reconstructed.
We could have also raised the cut on $E_{Tb}$ to $40$ or $50$ GeV, but
the invariant mass is somewhat more selective.

\section{Conclusions}

Cuts are virtually always made on angular distributions.  Some of
these cuts, like rejecting back-to-back jets to reduce
mismeasurements, are difficult to avoid.  Others enter indirectly from
kinematically-dependent efficiencies, or reconstruction cuts, like
fitting the top-quark mass.  These cuts generically cause initially
uncorrelated distributions, like $\cos\theta^t_{ej_1}$ for $Wjj$, to
appear to fake our signals.  By using more of the measured angles in
the event, these artificially induced correlations can be reduced or
removed.  This will become even more vital at the LHC, where Standard
Model backgrounds to new physics are large.  To use this information
reliably, however, will require studies that confirm that the
predictions of the angular distributions are stable against higher-order
radiation.

We have examined the relationship between the leading-order and
next-to-leading order predictions of angular correlations of leptons
and jets in the single-top-quark and $Wjj$ final states.  The good
news is that LO matrix elements are sufficient to capture the complete
angular correlations.  The uncertainties induced by using LO matrix
elements are less than 5\% of the total acceptance in all cases.
Hence, showering Monte Carlo generators that are fed with the
spin-dependent matrix elements may be used to reliably construct an
experimental analysis.  The only place to be careful is in $t$-channel
production, where matrix elements matched to NLO predictions are
required to obtain the correct kinematics \cite{Sullivan:2004ie}.
These matched samples are already available and in use by both the CDF
and D0 Collaborations.

This paper does not attempt to perform a fully simulated signal and
background study.  Nevertheless, if we apply the four angular cuts
listed in Sec.\ \ref{sec:otherang}, we can estimate that the
significance only improves slightly, but the signal to background
ratio $S/B$ improves by a factor of $1.5$.  Adding the invariant-mass
cut of Sec.\ \ref{sec:masscut} improves the significance by roughly
25\%, but, more importantly, improves $S/B$ by almost a factor of 3.
The complete set of cuts retains about 40\% of the signal, but reduce
the $Wjj$ background by a factor of 7.  These results should be
further optimized after full detector simulations, but already
represent the power inherent in using well-predicted angular
correlations.

\begin{acknowledgments}
Research at Fermi National Accelerator Laboratory was supported by the
U.~S.\ Department of Energy, High Energy Physics Division, under
contract No.\ DE-AC02-76CH03000.  Research in the High Energy Physics
Division at Argonne National Laboratory was supported by an FY2005
grant from the Argonne Theory Institute, and by the U.~S.\ Department
of Energy under Contract No.\ W-31-109-ENG-38.
\end{acknowledgments}


\begin{thebibliography}{99}

\bibitem{CDF}
CDF Collaboration, C.~I.~Ciobanu,
Int.\ J.\ Mod.\ Phys.\ A {\bf 16S1A}, 389 (2001);
CDF Collaboration, D.~Acosta {\it et al.},
Phys.\ Rev.\ D {\bf 65}, 091102 (2002);
CDF Collaboration, T.~Kikuchi, S.~K.~Wolinski, L.~Demortier, S.~Kim,
and P.~Savard, Int.\ J.\ Mod.\ Phys.\ A {\bf 16S1A}, 382 (2001);
CDF Collaboration, D.~Acosta {\it et al.},
Phys.\ Rev.\ D {\bf 69}, 052003 (2004).

\bibitem{D0}
D0 Collaboration, V.~M.~Abazov {\it et al.},
Phys.\ Lett.\ B {\bf 517}, 282 (2001);
D0 Collaboration, B.~Abbott {\it et al.},
Phys.\ Rev.\ D {\bf 63}, 031101 (2001);
D0 Collaboration, A.~P.~Heinson,
Int.\ J.\ Mod.\ Phys.\ A {\bf 16S1A}, 386 (2001).

\bibitem{Acosta:2004bs}
CDF Collaboration, D.~Acosta {\it et al.},
Phys.\ Rev.\ D {\bf 71}, 012005 (2005).

\bibitem{Abazov:2005zz}
D0 Collaboration, V.~Abazov {\it et al.},
Phys.\ Lett.\ B {\bf 622}, 265 (2005).

\bibitem{Bowen:2005rs}
M.~T.~Bowen, S.~D.~Ellis, and M.~J.~Strassler,
Acta Phys.\ Polon.\ B {\bf 36}, 271 (2005).

\bibitem{Cao:2004ap}
Q.~H.~Cao, R.~Schwienhorst, and C.~P.~Yuan,
Phys.\ Rev.\ D {\bf 71}, 054023 (2005); 
Q.~H.~Cao, R.~Schwienhorst, J.~A.~Benitez, R.~Brock, and C.~P.~Yuan,
hep-ph/0504230.

\bibitem{Stelzer:1998ni}
T.~Stelzer, Z.~Sullivan, and S.~Willenbrock,
Phys.\ Rev.\ D {\bf 58}, 094021 (1998).

\bibitem{Mahlon:1996pn}
G.~Mahlon and S.~Parke, Phys.\ Rev.\ D {\bf 55}, 7249 (1997);
Phys.\ Lett.\ B {\bf 476}, 323 (2000).

\bibitem{MCFM}
J.~M.~Campbell and R.~K.~Ellis, Phys.\ Rev.\ D {\bf 62}, 114012 (2000);
Phys.\ Rev.\ D {\bf 65}, 113007 (2002);
J.~Campbell, R.~K.~Ellis, and F.~Tramontano,
Phys.\ Rev.\ D {\bf 70}, 094012 (2004).

\bibitem{Sullivan:2004ie}
Z.~Sullivan,
Phys.\ Rev.\ D {\bf 70}, 114012 (2004).

\bibitem{Pumplin:2002vw}
J.~Pumplin, D.~R.~Stump, J.~Huston, H.~L.~Lai, P.~Nadolsky, and W.~K.~Tung,
J.\ High Energy Phys.\ {\bf 07}, 012 (2002).

\bibitem{Stump:2003yu}
D.~Stump, J.~Huston, J.~Pumplin, W.~K.~Tung, H.~L.~Lai, S.~Kuhlmann, and
J.~F.~Owens, J.\ High Energy Phys.\ {\bf 10}, 046 (2003).

\bibitem{Harris:2002md}
B.~W.~Harris, E.~Laenen, L.~Phaf, Z.~Sullivan, and S.~Weinzierl,
Phys.\ Rev.\ D {\bf 66}, 054024 (2002).

\bibitem{O'Neil:2002ks}
ATLAS Collaboration, D.~O'Neil {\it et al.},
J.\ Phys.\ G {\bf 28}, 2657 (2002).

\bibitem{Sjostrand:2000wi}
T.~Sj\"ostrand {\it et al.},
Comput.\ Phys.\ Commun.\  {\bf 135}, 238 (2001).

\bibitem{Corcella:2000bw}
G.~Corcella {\it et al.}, J.\ High Energy Phys.\ {\bf 01}, 010 (2001).

\end{thebibliography}
\end{document}